\documentclass{iau}
\usepackage{graphicx}
\usepackage{hyperref}
\hypersetup{colorlinks=true, breaklinks=true, citecolor=blue, linkcolor=blue, menucolor=blue, urlcolor=blue}

\newcommand{\so}{3C~454.3}
\newcommand{\mg}{Mg~II$\lambda 2800$\AA}
\newcommand{\cnt}{UV-continuum~$\lambda 3000$\AA}
\newcommand{\mm}{$\lambda 8$mm}

\title[Broad-line emission driven by  non-thermal continuum from the jet] %% [give here short title] %%
{The link between broad emission line fluctuations and non-thermal emission from the inner AGN jet}

\author[J. Le\'on Tavares et al. ]   %% [give here the short author list; use "et al." if 3 authors or more] %%
{
J. Le\'on Tavares$^1$,  
V. Chavushyan$^1$ 
A. Lobanov$^{2, 3}$
E. Valtaoja$^4$
\and
T.~G.~Arshakian$^5$
}

\affiliation{
$^1$ Instituto Nacional de Astrof\'{\i}sica \'Optica y Electr\'onica (INAOE), 
         Apartado Postal 51 y 216, 72000 
         Puebla, M\'exico \\
[\affilskip]
$^2$ Max-Planck-Institut f\"ur Radioastronomie, Auf dem H\"ugel 69, 53121 Bonn, Germany \\
[\affilskip]
$^3$ Institut f\"ur Experimentalphysik, Universit\"at Hamburg, Luruper Chaussee 149, 22761 Hamburg, Germany\\
[\affilskip]
$^4$  Tuorla Observatory, Department  of Physics and Astronomy, University of Turku, 20100 Turku, Finland\\
[\affilskip]
$^5$ I. Physikalisches Institut, Universit\"at zu K\"oln, Z\"ulpicher Str. 77, 50937 K\"oln, Germany\\
email: {\tt leon.tavares@inaoep.mx}
}

\pubyear{2014}
\volume{313} 
\pagerange{xx--xx}
\jname{Extragalactic jets from every angle}
\editors{F. Massaro, C.C. Cheung, E. Lopez, A. Siemiginowska, eds.}
\begin{document}

\maketitle

\begin{abstract}
AGN reverberate when the broad emission lines respond to changes of the ionizing thermal continuum emission. Reverberation measurements have been commonly used to estimate the size of the broad-line region (BLR) and  the mass of the central black hole.  However, reverberation mapping studies have been mostly performed on radio-quiet sources where the contribution of the jet can be neglected. In radio-loud AGN, jets and outflows may affect substantially the relation observed between the ionizing continuum and the line emission. To investigate this relation, we have conducted a series of multi-wavelength studies of radio-loud AGN, combining optical spectral line monitoring with regular VLBI observations. Our results suggest that at least a fraction of the broad-line emitting material can be located in a sub-relativistic outflow ionized by non-thermal continuum emission generated in the jet at large distances ($>$ 1~pc) from the central engine of AGN. This finding may have a strong impact on black hole mass estimates based on measured widths of the broad emission lines and on the gamma-ray emission mechanisms.

\keywords{galaxies: nuclei, galaxies: jets, quasars: emission lines, galaxies: individual(3C~454.3), galaxies: individual(3C~390.3),  galaxies: individual(3C~120),  gamma rays: observations.  }
\end{abstract}

\firstsection % if your document starts with a section,
\section{Introduction}

The correlated variability of the optical-UV continuum emission and the strength of the broad emission lines in AGN has been known and characterised with monitoring campaigns over  the last decades (e.g. \cite[Blandford \& McKee 1982]{blandford_1982}, \cite[Peterson 1993]{1993PASP..105..247P}, \cite[Shapovalova et 
al. 2001]{2001AA...376..775S}, \cite[Shapovalova et 
al. 2013]{2013A&A...559A..10S}). The  delay  between increases in the continuum optical-UV emission  and the onset of the increase in the intensity of the broad emission lines  allows to obtain  a direct estimation  about the distance  of the broad-emission line region to the ionizing source. Then, assuming that the BLR clouds  follow a virialized motion, the mass of the central black hole can be estimated. The  so-called reverberation mapping technique  has been used to determine structure and dynamics of the BLR and to provide scaling relations  to estimate black hole masses by characterizing single-epoch spectra (\cite[e.g. Vestergaard 
\& Peterson 2006 ]{2006ApJ...641..689V}).   \\

However, reverberation mapping studies have been traditionally performed  mostly on radio-quiet AGN where the dominant source of continuum emission  has a thermal origin (e.g. accretion disk). Nevertheless, in radio-loud sources, the non-thermal emission produced in the inner regions of  relativistic jets (see Figure~\ref{fig1}) can dominate  at all energies and could become an important source of ionizing continuum.  So the question arises: Can the non-thermal continuum emission from the inner jet ionize  clouds of the BLR?

\section{3C~454.3}

We have explored the variability of the broad emission lines in 3C 454.3 (so far, the brightest blazar seen by  the \emph{Fermi} gamma-ray space telescope) in order to use it as an auxiliary piece of information to probe the geometry and physics of its innermost regions and to discriminate between scenarios of the gamma-ray production.  The  optical spectra of \so\  have been acquired,  as part of  the  Ground-based Observational Support of the Fermi Gamma-ray Space Telescope at the University of Arizona monitoring program (\cite[Smith et al. 2009]{2009arXiv0912.3621S}),  over a period of three years (2008-2011) and due to its  redshift ($z=0.859$)  we have had access to the middle-UV region of the spectrum, allowing us to  monitor its \mg\ broad-emission line and  adjacent \cnt.  

\begin{figure}[t]
\begin{center}
\includegraphics[angle=90,width=1.1\columnwidth]{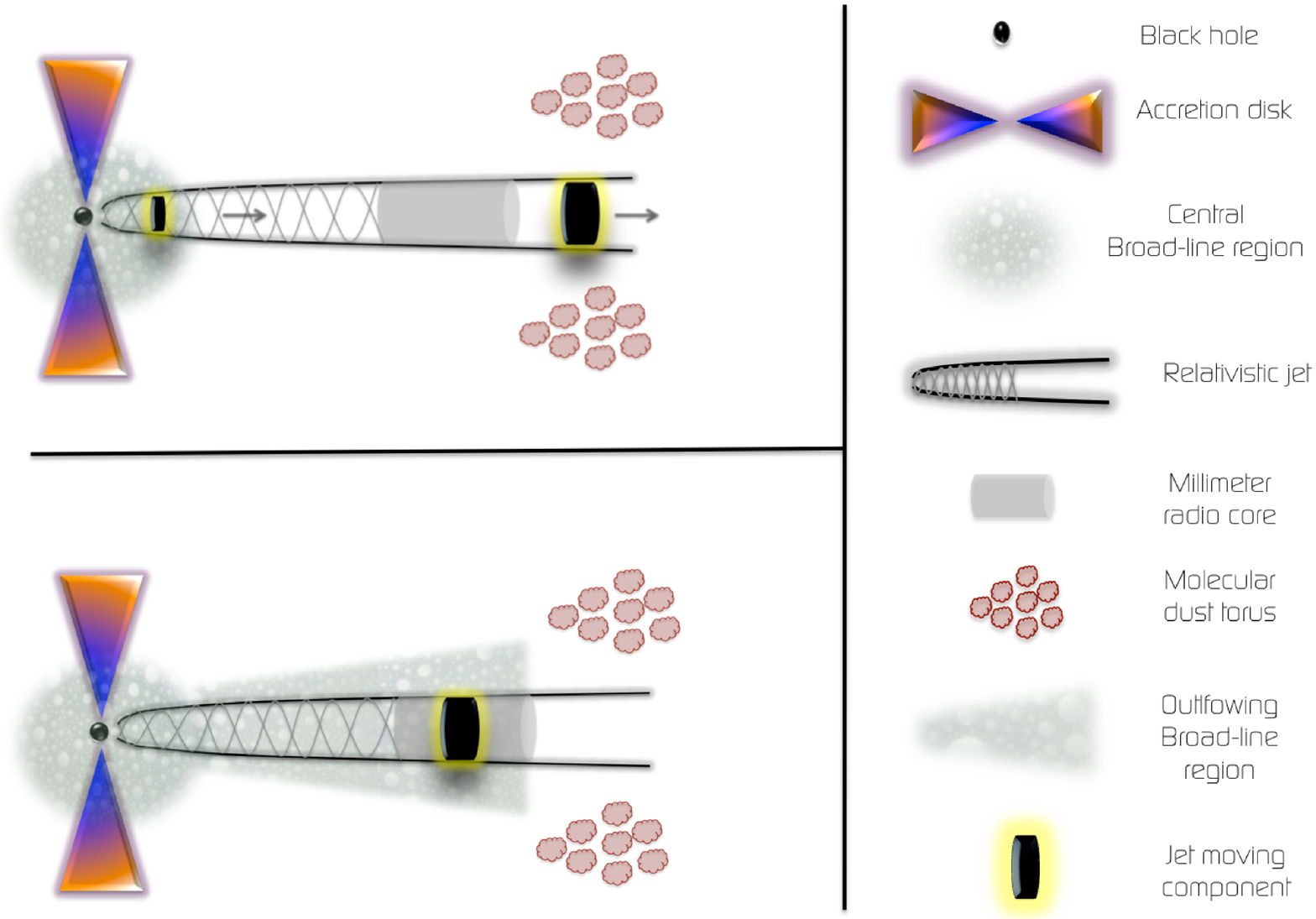}
\caption{\emph{Top-left panel}: Sketch of the inner regions of  a radio loud AGN. \emph{Right panel:}  Identification of each of the inner AGN components. \emph{Bottom-left panel:} Sketch of the inner regions of the $\gamma$-ray blazar 3C~454.3 (\cite[Le{\'o}n-Tavares et al. 2013]{2013ApJ...763L..36L}). The dimensions in the sketch are not in scale.  }\label{fig1}
   \label{fig1}
\end{center}
\end{figure}

\begin{figure}[t]
\begin{center}
\includegraphics[scale=0.5]{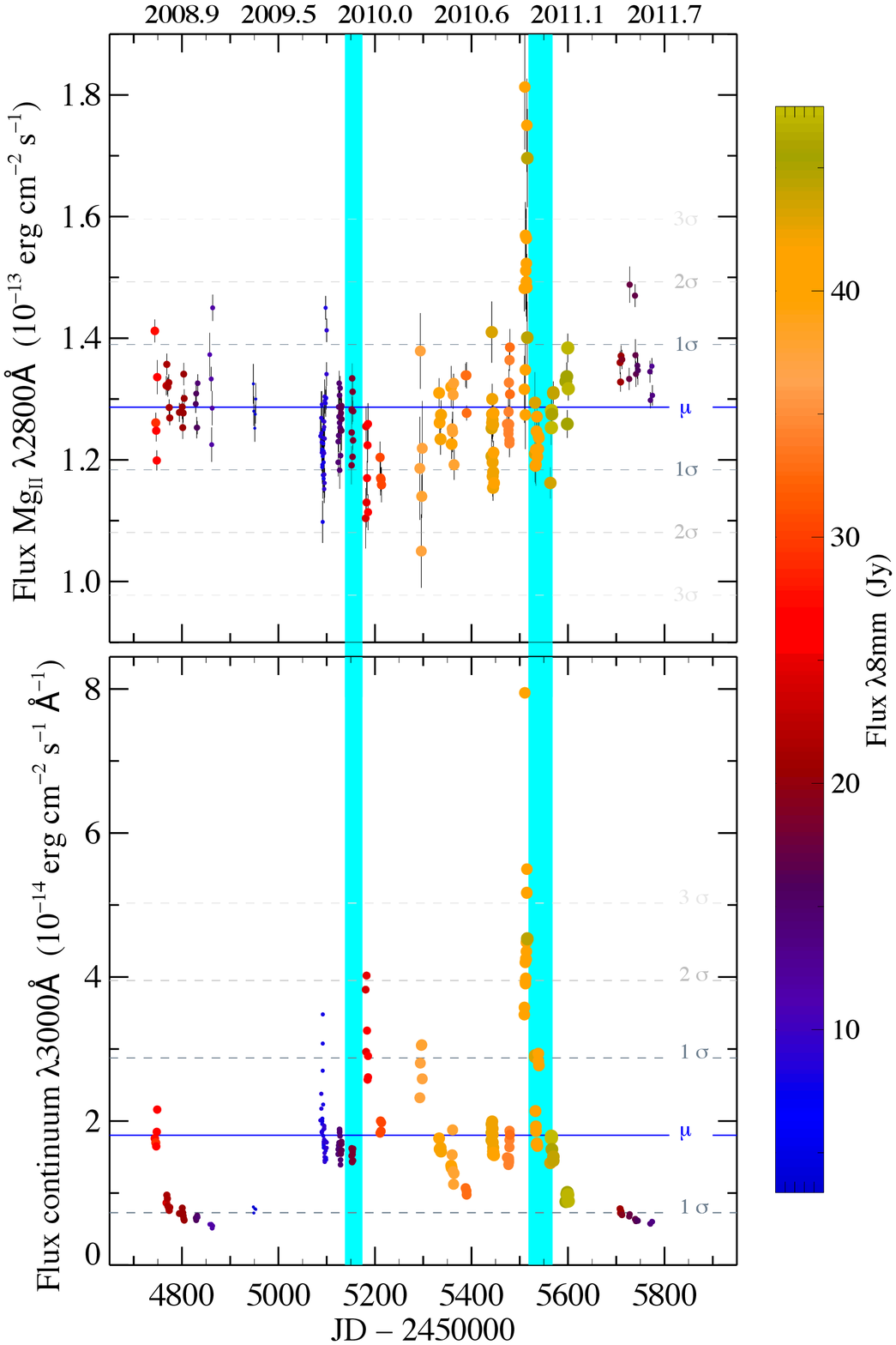}
\caption{Flux evolution of  \mg\  emission line (\emph{top panel}) and \cnt\  emission (\emph{bottom panel}). For each panel, the solid (blue) horizontal line denotes the mean flux ($\mu$)  observed during  the monitoring period, whereas  dashed (gray)  horizontal lines show multiples of $\sigma$, where $\sigma$ is the standard deviation of the flux. In this work, we consider a significant flare if the levels of emission exceed 2$\sigma$. Symbol size and color are  coded according to the color bar displayed, where the larger and lighter the symbols,   the higher the  level of  \mm\ emission observed. The vertical stripes  show the time when new blobs were ejected from the radio core and their widths  represent the associated uncertainties.  }\label{fig2}
\end{center}
\end{figure}

\newpage

In \cite[Le{\'o}n-Tavares et al.(2013)]{2013ApJ...763L..36L}, we found for the first time  a statistically significant  flare-like event  in the   \mg\ light curve of \so. As shown in Figure \ref{fig2}, the highest levels of \mg\ line flux ($\geq 2\sigma$) occurred after   \mm\ flare onset, during an increase in the optical polarization percentage,  before  the emergence of a new superluminal component from the radio core  and within the largest gamma-ray flare ever seen.   This finding crucially links the broad-emission line fluctuations  to  the non-thermal continuum emission   produced by relativistically moving  material in the  jet and hence to the presence of broad-line region clouds surrounding the radio core (see bottom-left panel of Figure~\ref{fig1}). The variability of other broad-emission lines in 3C~454.3   has been  reported in \cite[Isler et al.(2013)]{2013ApJ...779..100I}.

\begin{figure}[t]
\begin{center}
\includegraphics[angle=90,scale=0.45]{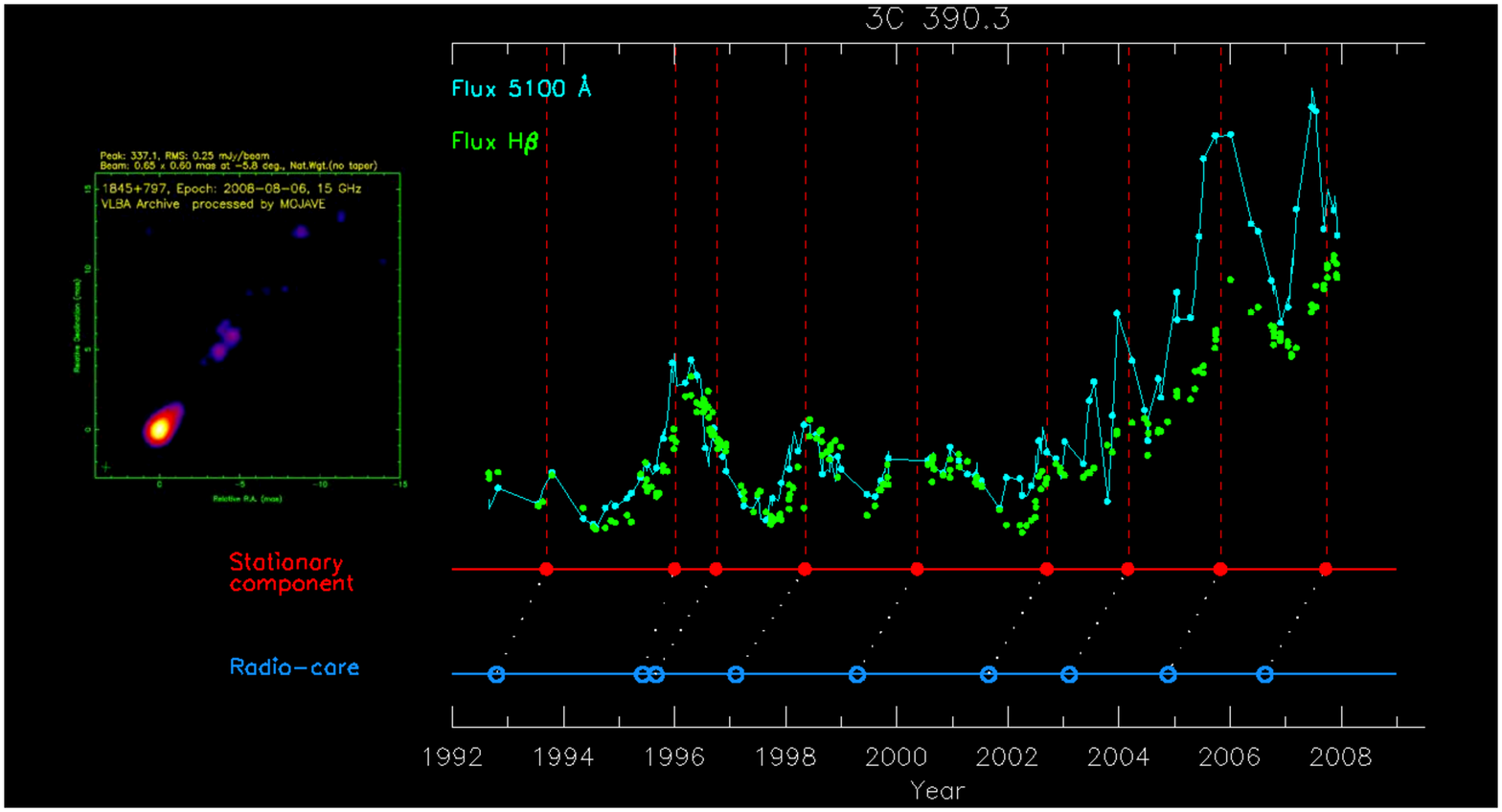}
\caption{Variability of the optical continuum(5100~\AA), H$\beta$ broad-line emission and jet kinematics in the radio galaxy  3C~390.3 (\cite[Arshakian et al. 2010]{2010MNRAS.401.1231A}).  All optical flares  (both continuum and H$\beta$ broad emission line) are associated with component jet ejection events: the flare rises after the epoch of ejection of a new jet component from the radio-core and it reaches the maximum around the epoch at which the ejected radio knot passes through a stationary   component  downstream the jet. The maximum of H$\beta$ broad-line emission occurs when a component passes through a stationary component in the inner jet  located at a distance of $\sim~0.5$~pc from the radio-core (the radio core might be located  at several parsecs from the black hole). This result strongly  suggests: (i)  that  the jet can power a significant amount of broad-line emission  and, (ii)  the presence of broad-line region material at distances well beyond the canonical BLR ($<$~1~pc).  }\label{fig3}
\end{center}
\end{figure}

\section{3C~390.3 and 3C~120}
Arshakian et al. (2010) and Le\'on-Tavares et al. (2010) found that for the radio galaxies 3C 390.3\footnote{\href{http://www.metsahovi.fi/~leon/movies/3c3903.gif}{http://www.metsahovi.fi/$\sim$leon/movies/3c3903.gif}} and 3C 120\footnote{\href{http://www.metsahovi.fi/~leon/movies/3c120.gif}{http://www.metsahovi.fi/$\sim$leon/movies/3c120.gif}}, the variable optical continuum starts to rise when a new superluminal component leaves the radio core seen at 15 GHz and its maximum occurs when the component passes through a stationary feature located downstream of the radio core. Since these two radio galaxies are known to reverberate (Shapovalova et al. 2010; Grier et al. 2012), in the sense that the H$\beta$ broad emission line responds to changes in the optical continuum,  the authors conclude that the jet can power a significant amount of broad-line emission particularly during strong  non-thermal continuum flares  from the jet (see also \cite[Belokon' 1987 ]{1987Ap.....27..588B} and  \cite[Perez et al. 1989]{1989MNRAS.239...75P}).

\section{Summary}

The above  results suggest the presence of an additional component of the BLR,  dubbed as \emph{outflowing BLR} (see bottom-left panel of Figure \ref{fig1}),  which in effect might be filled with BLR material dragged by the relativistic jet as it propagates downstream of the black hole or perhaps  could be a sub-relativistic outflow arising from an accretion-disk  wind.   The presence of broad-line region material surrounding  and being ionized by the radio core  (at distances beyond the inner parsec)   have far reaching implications for the current AGN energy release models (\cite[Le{\'o}n-Tavares et 
al. 2011b]{2011A&A...532A.146L})   as well as for black hole mass estimates made in AGN using the technique of reverberation mapping and its scaling relations. 

The fact that broad emission lines in radio-loud AGN might respond to changes of the non-thermal continuum prevents us from using the single epoch virial black hole mass estimates because the latter assumes: (1) a single localized ionization source (i.e., accretion disk) and (2) virial equilibrium of the BLR clouds. The latter assumptions cannot be fulfilled during episodes of strong flaring activity, hence the ionization of BLR clouds by non-thermal emission  from the jet might introduce uncertainties to the black hole mass estimates derived by assuming virial equilibrium of the BLR. Alternative scaling relations to weigh the black holes  in strongly beamed sources are discussed and implemented in \cite[Le{\'o}n-Tavares et al.(2011a, 2014)]{2011MNRAS.411.1127L, 2014ApJ...795...58L}.

\end{document}